# Stray light analysis of the Xinglong 2.16-m telescope

Taoran Li[a]

[a]Key Laboratory of Optical Astronomy, National Astronomical Observatories Chinese Academy of Sciences, 100101, Beijing, China;

**ABSTRACT**

An effort towards understanding of the stray light problems for the Xinglong 2.16-m telescope was presented to estimate the stray light performance of the telescope itself and provide a method for improving the stray light suppression. The stray light analysis for 2.16-m telescope model, which consists the onion shaped dome, telescope structure, equatorial mount and telescope optics, has been performed with two cases (1) point to 60° and (2) point to zenith, in both azimuth and elevation. A simplified mechanical model obtained from the detailed telescope system geometry was used to increase the simulation efficiency. The source setting, property assumptions and ray tracing methodology for stray light simulation are described. The critical and illuminated objects were identified to understand the stray light propagation path in system. The Point source normalized irradiance transmittance (PSNIT), which is generally used for assessing stray light and uncorrelated to entrance aperture was calculated utilizing the stray light software TracePro to evaluate the stray light performance of the 2.16-m telescope with a series of off-axis angles. It shows that the PSNIT values are less than $10^{-7}$ when off-axis angles are larger than ±20°. The dominant objects for stray light (primary and secondary mirror, telescope structure and dome) were identified to give advice for performance improvement. The analyses indicate that significant benefit can be realized with adding only 5 vanes inside the bottom portion of the secondary baffle. In the case of point to zenith, the PSNIT values will decrease about 40% at average.

**Keywords:** Xinglong 2.16-m telescope, Stray light analysis, TracePro, PSNIT, Vanes

## 1. INTRODUCTION

The Xinglong 2.16-m telescope, inaugurated in 1989, located at the Xinglong Observatory, is a Ritchey-Chretien (R-C) telescope, with a primary mirror of diameter 2.16 meters[1]. The Xinglong 2.16-m telescope used to be the largest optical telescope in China and even in the Far East, it is a symbol of our country being able to research and develop large precision equipment independently.

The optical system of 2.16-m telescope is represented on figure 1. The hyperboloidal primary mirror M1, is made of a JIK5 glass from Soviet Union, concentrating 80% of the light into an area 0.6 arcsec in diameter. A convex secondary mirror M2, made of Zerodur, 720mm in diameter, together with the primary make the effective focal length for the Cassegrain focus 19.4m (f /9) [2]. The telescope also incorporated a tertiary mirror (not shown in figure 1), which can divert the light from M2 to the Coudé focus (retired since 2013). As shown in figure 2, the 2.16-m telescope has the tube of conventional open truss form and the mounting of the equatorial type. The initial opto-mechanical design comprises 2 baffles, which are M2 baffle surround the M2 and the primary baffle above the M1. The Cassegrain common interface including the calibration device and guiding device provides the calibration lamps and autoguider systems, and a mounting for the Cassegrain instruments.

During the observations performed over 30 years, a few baffling solutions (vanes on inner surface of primary baffle) and some tests have been made, but the stray light performance of the telescope remains unknown, which is critical for the high-performance requirement of the optical instruments. Sometimes unidentified scattered light could be seen on the detector on bright night.

This paper describes efforts towards an understanding of the stray light problems with the telescope itself without the instruments. The stray light analysis is aimed to estimate the stray light performance of the 2.16-m telescope and provide an overview of the stray light suppression for future instruments (e.g., a new intermediate-resolution spectrograph and Tip-tilt module). The analyses are based on ray tracing model that includes the dome, telescope structure and optics. The software SOLIDWORKS, TracePro and MATLAB are used for modeling, ray tracing and analyzing, respectively.

The paper is organized as follows: Section 2 shows the stray light model of 2.16-m telescope, definition of Point source normalized irradiance transmittance and simulation settings; Identification of critical objects and illuminated objects are described in Section 3; The analyses of PSNIT curve and stray light contributors in two cases are presented in Section 4 and 5; Finally, a brief conclusion is discussed in Section 6.

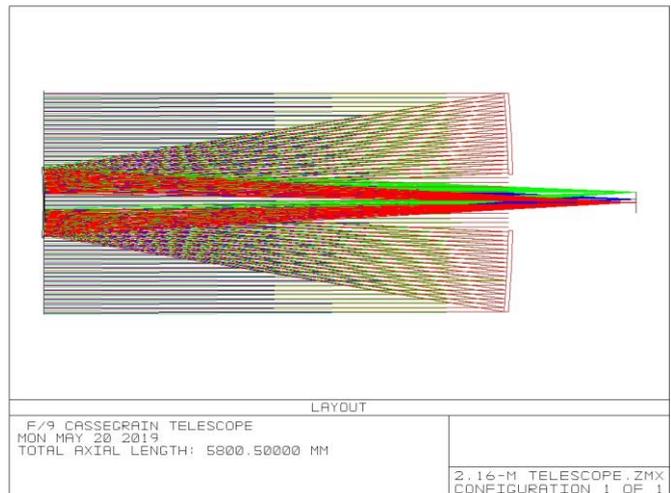

Figure 1 Optical layout of the Xinglong 2.16-m telescope 's f/9 Cassegrain focus. The diameter of the primary mirror is 2.16m. Red rays are from an axis point source at infinity. Blue and green rays are from a source at radius 6 arcmin and 18 arcmin, respectively.

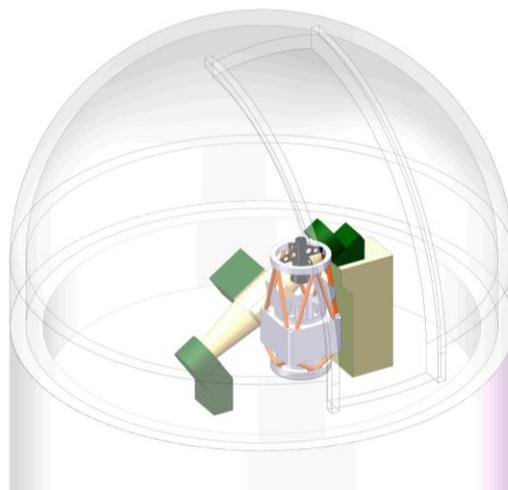

Figure 2 SOLIDWORKS model of the Xinglong 2.16-m telescope, including the enclosure (rotating dome with slit, in transparent), equatorial mount (south and north pier, in emerald green and dark green), mirror cell, Serrurier truss, top-end ring, cylindrical baffles for the primary (M1) and secondary (M2) mirrors, vanes on inner surface primary baffle.

## 2. STRAY LIGHT MODEL OF XINGLONG 2.16-M TELESCOPE

### 2.1 Stray light analysis

Stray light is generally defined as unwanted light that reaches the focal plane of an optical system[3]. It is one of the most important factors that influence the high-accuracy photometry of optical astronomical telescopes, which could reduce the signal to noise ratio of astronomical objects. Stray light analysis begins with the 3D modeling of the optical and mechanical geometry. Then, rays tracing is ready as the simulation parameters settings and the surface property assignments are completed in ray tracing software (TracePro).

### 2.2 Telescope and dome simplification

The original model contains tiny parts and complex shapes which can increase simulation time and are not necessary. Therefore, the mechanical model should be simplified (e.g., removing the screws and reconstructing the top end assembly as an annular part).

The simplified TracePro model has 8 basic components:
(1). Dome
(2). Mount
(3). Telescope structure (spider, top end assembly, focus drive, truss, center section, cassegrain focus interface)
(4). Secondary baffle
(5). Primary baffle
(6). Vanes (on inner surface of primary baffle)
(7). Optical system
(8). CCD

The 2.16-m telescope has an onion shaped dome, of 22 meters internal diameter, and an up-and-over shutter (5200mm wide). The distance between the top of the slit and the top end of the telescope is about 8700mm. The inner diameter of the primary and secondary baffle is 574mm and 960mm. Sketching additional rays representing moonlight varied as azimuth show that the scattered light will be blocked from directly reaching (1) the primary baffle by the edge of secondary baffle, and (2) the M1 by the edge of secondary baffle and dome slit, if the angular distance from the moon is less than ±7.1° and larger than ±13.2°, respectively (positive: right side of the dome slit). As in elevation, the moonlight will be blocked by (1) the edge of dome slit and secondary baffle, and (2) center section, if the moon distance is larger than +14.6° and -95.6°, respectively (positive: moon is higher than telescope), as shown in figure 3.

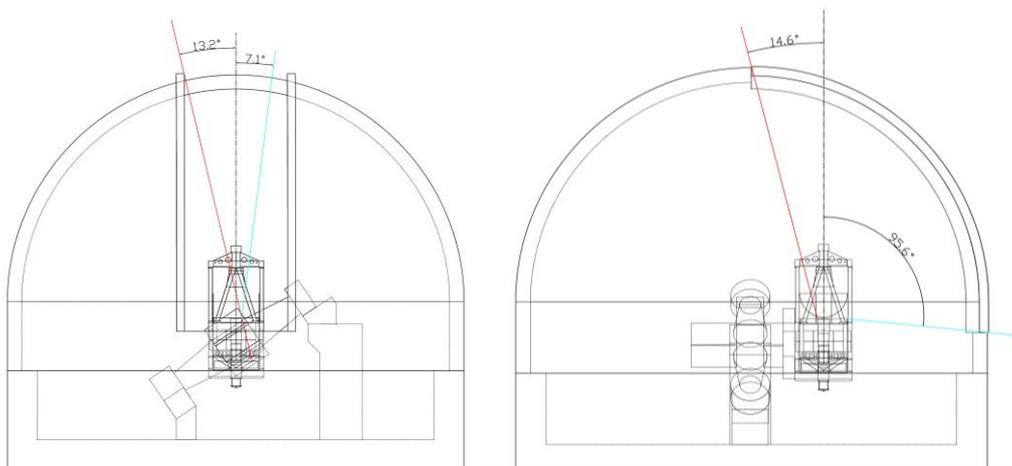

Figure 3 Layout of dome and simplified 2.16-m telescope components in profile view, some limiting lines of sight from the moon to different points on the telescope structure. (a) Stray light source varied as azimuth, a view through the dome slit, showing that light from the moon cannot enter the M1 baffle when the moon's angular distance from the telescope pointing is < 7.1° (blue line, blocking by the M2 baffle) or > 13.2° (red line, blocking by edge of dome slit); (b) Stray light source varied as elevation, a view with the dome rotated by 90°, dome slit at right, showing that light from the moon cannot enter the M1 baffle if the moon's zenith distance is more than 14.6° in the direction away from the dome slit (red line, blocked by edge of dome slit ) and it cannot impinge on M1 directly if the zenith distance is more than 95.6° (blue line, blocked by telescope center section).

## 2.3 Point source normalized irradiance transmittance (PSNIT)

PSNIT (Point source normalized irradiance transmittance) is a transfer function which describes the stray light performance in an optical system, equals to the average irradiance over the focal plane divided by the incident irradiance at the instrument[4]. PSNIT is one of the types of PST (Point Source Transmittance), used in some systems that the entrance aperture may not be well defined. Due to the open-truss structure used at 2.16-m telescope, it is hard to define the entrance aperture. Here the PSNIT is calculated by,

$$PSNIT = \frac{E_{focal-plane}}{E_{incident}} = \frac{P_{focal-plane}/area_{focal-plane}}{P_{incident}/area_{incident}} , \qquad (1)$$

where $E$ is the irradiance defined as the total power $P$ per area. The irradiance of incident source is set to $1\,W/m^2$, and the area of the focal plane is $30mm \times 30mm$, therefore PSNIT is relevant only to the power on the focal plane. The source beam is uniform spatial and angular profile, entering the system randomly. The total rays are 1 million.

### 2.4 Source settings

The stray light sources are generally viewed as the point sources. The light sources for simulation are depended on the off-axis angle of the moon. The size of each source should be large enough to cover the whole dome shutter, which is the only path of the moonlight coming into the dome. The sources used for azimuth (left-right) and elevation (up-down) simulation are in the plane determined by the optical axis and the altitude of the telescope, perpendicular and parallel to the opening of the dome shutter, respectively (figure 4). The stray light sources in both azimuth and elevation range from 0° to ±90°. Each source has one million rays.

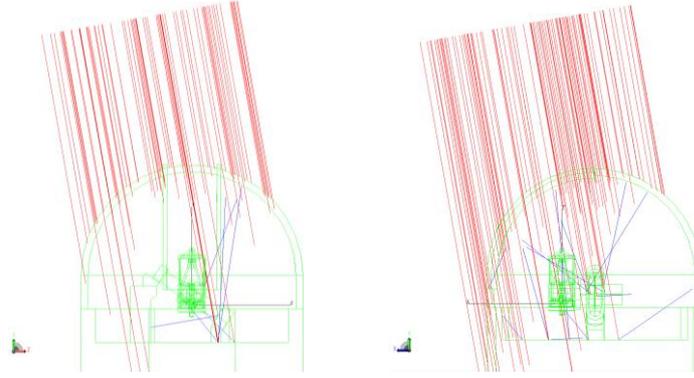

Figure 4 Azimuth and elevation simulation (in TracePro) of light from a point source at infinity, impinging on the dome, telescope and support structure. The telescope is pointed to zenith. Left: azimuth; right: elevation. Only for showing the location of the sources, not the same as the following simulation.

### 2.5 Surface property

BSDF (Bidirectional Scattering Distribution Function) is a common method used to describe the scattering characteristic of a surface. BRDF (Bidirectional reflectance distribution function) and BTDF (Bidirectional transmittance distribution function) are subclass of BSDF, used to describe the reflectance and transmittance, respectively[5].

The BSDF is a function of many parameters, normally it is difficult to develop highly accurate theoretical models. However, due to the time and budget required, the precise surface property is hard to measure. Fortunately, there are many surface properties useful both empirically and theoretically in TracePro. It refers to absorptance, specular reflectance, transmittance, BRDF and BTDF. As the specular reflectance is easy to obtain, the optical elements (M1 and M2) use the default "mirror" model in database, no dust on the optical surface, and just modify the specular reflectance as the measured value. The rest surfaces are treated as "black paint" to absorb the stray light at utmost. Table 1 shows the surfaces and their parameters.

Table 1 Surface property parameters in TracePro

| | Surface | Property |
|---|---|---|
| 1. | Mirror surface of M1 and M2 | Absorptance:9.9%, Specular reflectance:90.0%, Integrated BRDF:0.1% |
| 2. | Other surfaces | Absorptance:90.0%, Specular reflectance:2.0%, Integrated BRDF:8.0% |
| 3. | Outside surfaces of the dome | Perfect absorber-Absorptance:100.0% |

### 2.6 Importance sampling

Importance sampling is a technique used with Monte Carlo calculations to improve efficiency[6]. The "important" surfaces are forced to occur with higher probability than their nature. The stray light model with importance sampling could avoid using millions of the rays and perform the ray tracing in a reasonable time. An "importance" surface should have one or more targets where the rays supposed to go, either toward or away from the "importance" surface. In stray light simulation, the importance sampling should be selected on each first-order scattering surfaces.

## 3. CRITICAL AND ILLUMINATED SURFACES

The first step of ray tracing is to identify the critical and illuminated surfaces. A critical surface is one that can be seen by the detector or reflected by mirrors of the telescope. Mostly, all optical surfaces in telescope are critical surface. An illuminated surface is one that is illuminated by a stray light source directly or reflected by mirrors[7].

The first-order stray light path always occurs on a first-order scattering surface which is both critical and illuminated. The first-order paths are the most contributors of stray light[8]. In order to identify the critical and illuminated surfaces, the backward ray tracing from detector and the initial ray tracing from stray light source should be run.

The basic idea about the critical surfaces identification is to (1) analyze the system from the detector, (2) check the ray tracing history of every surfaces, and (3) find the surfaces visible from the detector. The initial ray tracing from stray light source is similar to the backward ray tracing. It utilizes several sources with different off-axis angles to find the surfaces which are illuminated directly. The critical and illuminated surfaces include the following:

**Critical surfaces:**
1. Cassegrain focus interface: inner surfaces
2. Center section: inner surfaces
3. M1: mirror surface
4. M2: mirror surface
5. Primary baffle: inner surface and upper edge
6. Secondary baffle: inner surface and bottom edge
7. Vanes: knife surfaces, upper and bottom surfaces

**Illumination surfaces:**
1. Primary baffle: inner surface, outside surface and upper edge
2. Secondary baffle: inner surface, outside surface and bottom edge
3. M1: mirror surface
4. M2: mirror surface
5. Vanes: knife surfaces and upper surfaces
6. Center section: inner surfaces
7. Other telescope structures (spider, top end assembly, truss)
8. Dome surfaces

The first-order scattering surfaces and their importance sampling targets are shown in Table 2.

Table 2  The first-order scattering surfaces and the importance sampling targets

| | **First-order scattering surface** | **Importance sampling target** |
|---|---|---|
| 1. | M1: mirror surface | Primary focus |
| 2. | M2: mirror surface | Focal plane |
| 3. | Primary baffle: inner surface and upper edge | Focal plane+ image of focal plane produced by M2 |
| 4. | Secondary baffle: inner surface and bottom edge | Focal plane |
| 5. | Center section: inner surfaces | Image of focal plane produced by M2 |
| 6. | Vanes: knife surfaces and upper surfaces | Image of focal plane produced by M2 |

## 4. POINT TO EL=60

The first attempt is pointing the telescope to 60° in elevation, as shown in figure 5. The dome shutter is corresponding with the telescope. An irradiance map (figure 6) shows the distribution and intensity of stray light on focal plane. Most of the stray light locate in a rectangular box on the focal plane due to the importance sampling. The total power in function (1) is derived from the total flux in irradiance map when simulations of each angles are finished. Figure 7 shows the plots of the PSNIT curves for azimuth (red line) and elevation (blue line) with logarithmic coordinates. The labels on the two plots illustrate the most significant contributors to the stray light flux at each angle. This is done by the "*Ray Path Sorting*" in TracePro (Table 3). The label "Tel.structure" includes the mount, spider, top end assembly, truss, center section and other parts of the structure of the telescope.

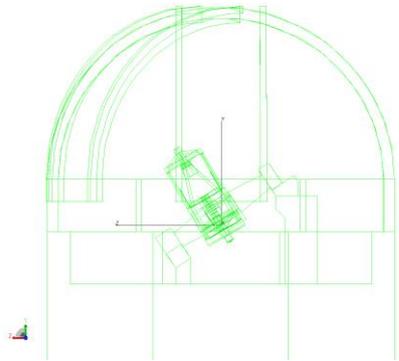

Figure 5 Model of telescope pointing to 60° in elevation. Due to the equatorial mount, the dome shutter is corresponding with the telescope.

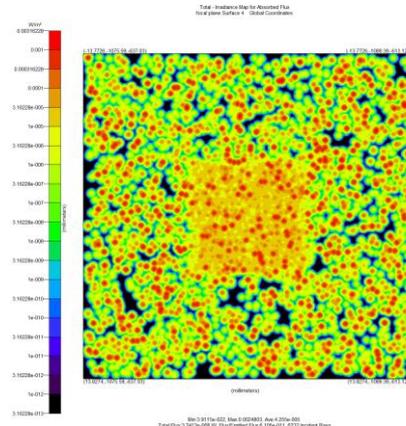

Figure 6 Irradiance map (predicted distribution of stray light) on focal plane for azimuth angle=5°, in watts per area. Most of the stray light locate in a box due to the importance sampling.

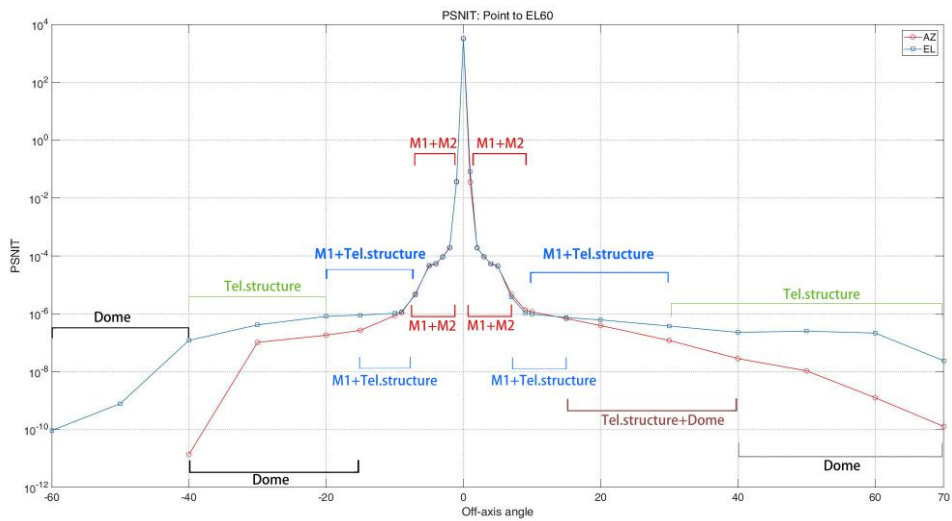

Figure 7 PSNIT curve for azimuth and elevation, telescope is pointed to 60°. Predicted relative intensity (PSNIT - see Section 2.3) of scattered moonlight, as a function angular distance of the moon in elevation (blue) and azimuth (red), when the telescope is pointed to elevation 60°. The dominant sources of scattered light are indicated by the labels above (for the elevation curve) and below (for the azimuth curve) the plotted curves. E.g. 'M1+M2' means that a large fraction of the scattered light arises from moonlight arriving directly on M1 but scattered up to M2 and then scattered down through the Cassegrain hole to the focal plane. The dramatic drops in scattered light intensity at azimuth beyond ± 7° are due to blocking of the moonlight by the secondary baffle (see Fig. 3). The blue line stops at 40° because no scatter light arrives on focal plane at azimuth angles > -40°.

Table 3 Ray Path Sort for 10° (ignore the ray paths that the percentage is less than 3%)

| Ray Path | No.of Rays | % of Total | Path Type | Object |
|---|---|---|---|---|
| 1 | 1618 | 42.49 | Single Scatter | Source-M1-M2-focal plane |
| 2 | 11 | 7.23 | Multiple Scatter | Source-Center section-Secondary baffle-focal plane |
| 3 | 16 | 4.34 | Multiple Scatter | Source-Mount-Secondary baffle-focal plane |
| 4 | 128 | 3.41 | Multiple Scatter | Source- Dome-M1-M2-focal plane |

### 4.1 Azimuth simulation

In azimuth simulation, the positive angles are located at the right side of the dome shutter (+z direction in figure 4 left). For the angles between the edge of the field of view (FOV) and ±10°, the PSNIT curve is nearly symmetrical with Y axis. The rays specular reflected by M1 and M2 dominate the stray light at the off-axis angles less than ±7°. The dramatic drops in scattered light intensity at azimuth beyond ± 7° are due to blocking of the moonlight by the secondary baffle. The structure of telescope is another contributor for the angles between ± 7°~± 15°, the rays can directly illuminate the outside surface of center section and the mount, then random scattering into the secondary baffle. M1 is not the contributor from ±15°, at which point the rays entering the dome shutter no longer directly illuminate M1 (figure 8 left). Beyond ±15°, the rays are gradually blocked by dome, the telescope structure and the inner surface of the dome contribute the most. As the source angle increase, the rays entering the dome move higher up the dome walls. For input angles greater than +40°, the inner surfaces of the dome are the most significant contributors.

Due to the English mount of 2.16-m telescope, the whole model is asymmetric. The PSNIT curve stops at -40° because no scatter light on focal plane at azimuth angles $\geqslant$ -40°. The PSNIT value is less than $10^{-7}$ when the off-axis angle is larger than 20°. The minimum PSNIT is $1.4\times10^{-11}$, at which angle (-40°) only one ray is scattered on the focal plane by random scattering of the dome surface.

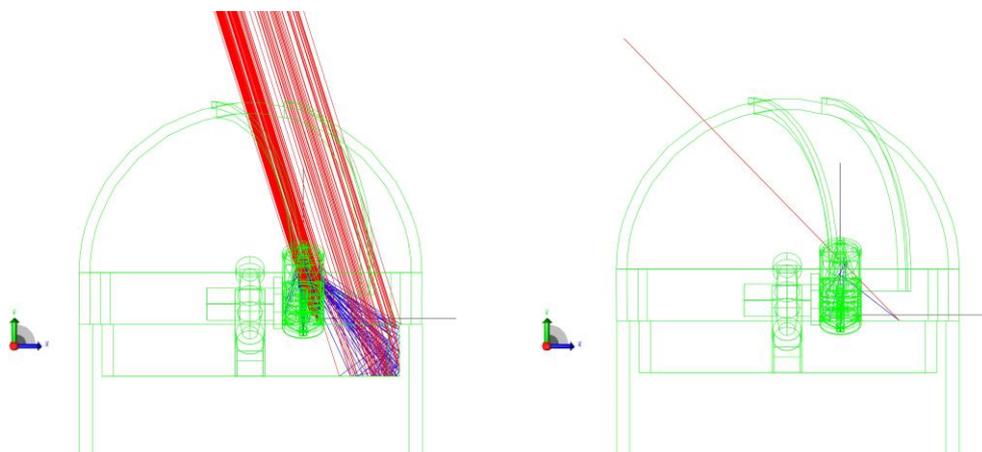

Figure 8 Ray tracing at -15° (left) and -40° (right) in azimuth simulation.

### 4.2 Elevation simulation

In elevation simulation, the positive angles are defined as the sources lower than the optics axis of the telescope. For the angles between 0° and ±10°, the PSNIT curves and the contributors of azimuth and elevation are almost the same. It's clear that the PSNIT values of elevation are larger than azimuth's after ±15°, because of the telescope structure scattering.

In positive direction, no dome will block the stray light from illuminating the telescope. M1 and telescope structure are the dominant sources from +15° to +30°. For elevation angles > +30°, the lights reflected from M1 no longer reach the M2 mirror cell, the telescope structure is the only contributor. The PSNIT curve drops gradually as the elevation angles fall toward horizon (the off-axis angles=60°), as shown in figure 9a.

For the curve along negative direction, it is almost symmetrical with the positive axis until the dome blocks the rays entering the telescope around -40° (figure 9b), the dome walls dominate the most scattered lights.

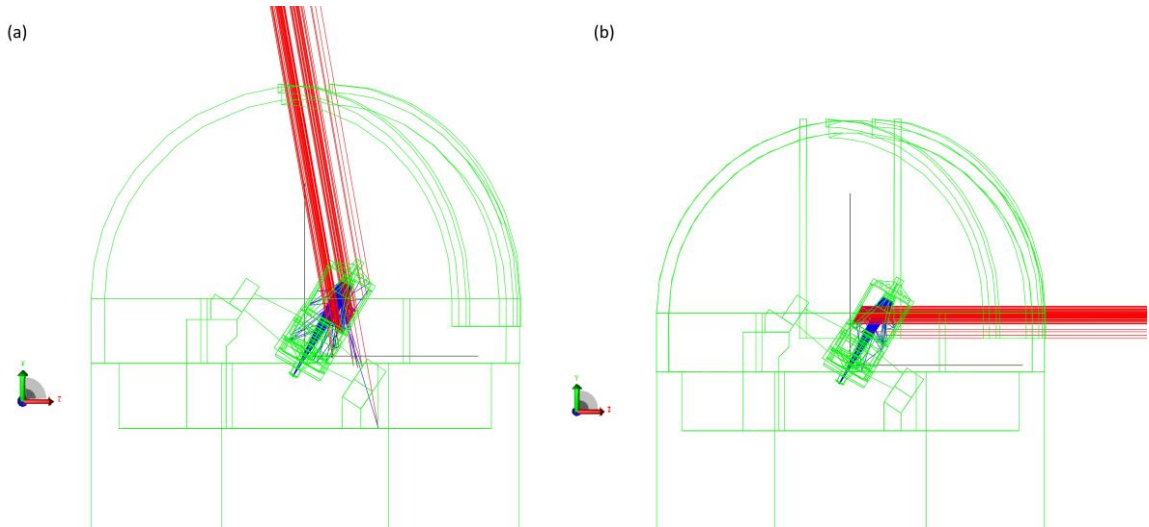

Figure 9 Ray tracing at +60° (a) and -40° (b) in elevation

### 4.3 Stray light distribution on focal plane

The TracePro software will consume almost all the computer memory when running a simulation. It's better to archive results from several simulations rather than one-time ray tracing. Therefore, the stray light distribution can be analyzed roughly by combining all the results at different angles.

Table 4 shows the ray numbers and fluxes on focal plane in two axes of azimuth (0~±40°) and elevation (0~±40°). For azimuth angles, the ray number has similar value in both sides of 0°, as well as the total flux. In positive axis of elevation, the flux is larger than that in negative axis because of no dome blocking.

Table 4 Ray number and flux on focal plane

|  | Azimuth | | Elevation | |
| --- | --- | --- | --- | --- |
|  | Ray number | Flux | Ray number | Flux |
| Positive axis | 59628 | 2.64e-05 | 60787 | 6.08e-05 |
| Negative axis | 57589 | 2.70e-05 | 62244 | 2.64e-05 |
| Total | 117217 | 5.34e-05 | 123031 | 8.72e-05 |

## 5. POINT TO ZENITH

Another attempt is pointing the telescope to zenith. In this case, the system is approximately symmetrical in azimuth direction. The max azimuth and elevation angles are 80° and 90°, respectively, as shown in figure 10.

### 5.1 Azimuth

From the edge of the FOV to ±10°, the rays can reach the focal plane by M1 and M2 reflection or scattering. The M1 and M2 are the most contributors. The PSNIT curve drops very fast as the off-axis angles increasing. No specular reflection by M1 happens at ±10°, where the PSNIT curve of azimuth has two steep drops. For the angle between ±10° and ±20°, the rays scattered by the M1, together with the rays scattered by the structure of the telescope dominate the total flux on focal plane. As the stray light source is far away from the dome slit (>20°), the rays can't illuminate the telescope directly, the dome (by random scattering) becomes the most contributor. Due to the several times random scattering by the dome, the curve is not smooth.

The PSNIT value is less than $10^{-8}$ when the off-axis angle is larger than 20°. The minimum PSNIT is $6.32 \times 10^{-11}$ at -80°. Compare with the azimuth PSNIT curve of the cases "point to 60°", the order of magnitude is similar with the case "point to zenith".

## 5.2 Elevation

The PSNIT curve of elevation simulation is nearly symmetrical with Y axis within ±15°. M1 and M2 are still the contributor for the elevation angle less than ±10°, the same as azimuth simulation.

As the elevation angle increase to +30°, rays can be scattered by M1 and the structure of the telescope. The PSNIT decreases slowly from +30° due to the random scattering by the telescope, the PSNIT value of elevation is about 1~2 orders of magnitude higher than the curve of azimuth simulation. The input rays are perpendicular to the telescope at +90°, only the rays scattered by dome will enter the light path.

For the elevation angle between -10° and -15°, the rays scattered by the M1 and the structure of the telescope dominate the total flux on focal plane. The dome shutter will block the rays from entering the telescope directly when elevation angle is larger than -15°, the dome surface becomes the stray light contributor. Beyond -40°, the stray light source comes to the back of the dome, all rays will be blocked.

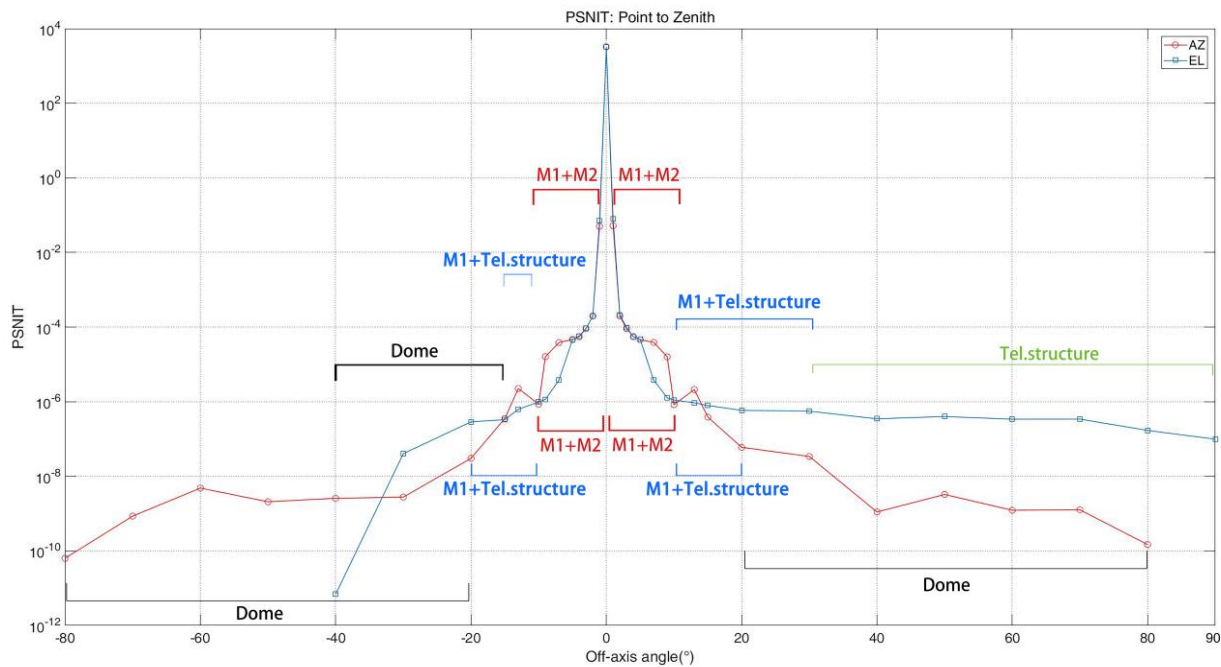

Figure 10 As Fig.7, PSNIT curve for azimuth and elevation, telescope is pointed to zenith.

## 5.3 Adding vanes inside the secondary baffle

In optical astronomical telescopes, the vanes are usually located inside the primary baffle and the secondary baffle, blocking the scattered light and improving the stray light performance.

As a result of analyzing the "*Ray Path Sorting*" table for all angles, it's clear that almost all the rays coming from the telescope structure (center section, mount ...) and dome will be scattered by the secondary baffle (M2 baffle) before arriving at the focal plane. Figure 11 shows the ray path for elevation angle 30° of the case "point to zenith". Considering about the above conclusions, adding vanes inside M2 baffle is a reasonable method to control the stray light.

In figure 12, the half profile of the M2 baffle is shown (solid line), with the marginal ray paths of optical system indicated as dot line. For simplicity, this is shown not to scale and without optical components, only the M2 baffle housing wall, the marginal rays are defined.

Based on the structure of the M2 baffle and the experiences of stray light control (the bottom part of the M2 baffle will block the most scattered light), only 5 vanes perpendicular to optical axis are used. The 5 vanes locate at inner surface of M2 baffle every 100mm from the bottom. The depth of each vane is 14.5mm in order to avoid marginal rays blocking.

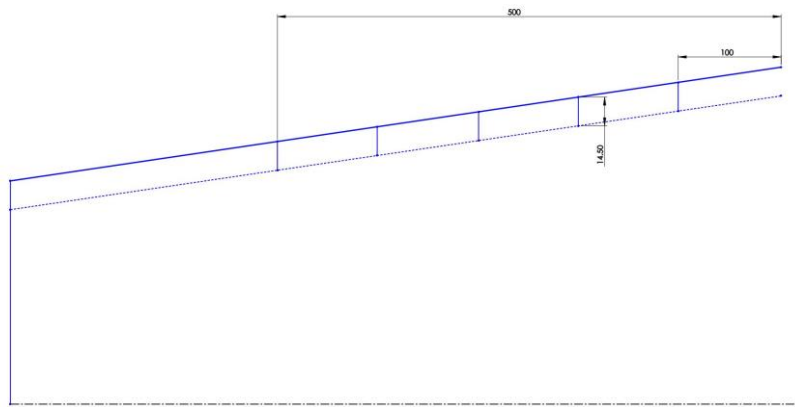

Figure 11 The "Ray Path Sorting" table for elevation angle 30° of the case "point to zenith", which shows the rays number, flux percentage of each ray, incident flux, path type and the ray path. "SpecRefl" means the specular reflection, "ImpRefl" means the importance sampling reflection/ scattering and "RandRefl" means the random reflection/scattering.

Figure 12 Vanes design for the M2 baffle of Xinglong 2.16-m telescope (not to scale). The solid and dotted line represent the profile of M2 baffle and the marginal rays, respectively. The dash-dotted line is the optical axis. The 5 vanes locate at inner surface of M2 baffle every 100mm from the bottom (right side). The depth of each vane is 14.5mm in order to avoid marginal rays blocking.

Figure 13 shows the PSNIT results with or without vanes in azimuth and elevation when telescope is pointed to zenith. It's clear that the vanes can improve the stray light performance in most cases, especially for the positive off-axis angle in both azimuth and elevation simulations. The downtrend of curves is not changed. The PSNIT values decrease about 40% at average.

The total number of rays is also reduced. As shown in Table 5, the total numbers under 14 off-axis angles for azimuth and elevation simulations decrease by approximate 8 thousand and 10 thousand, respectively. In elevation simulation, the suppression is more effective.

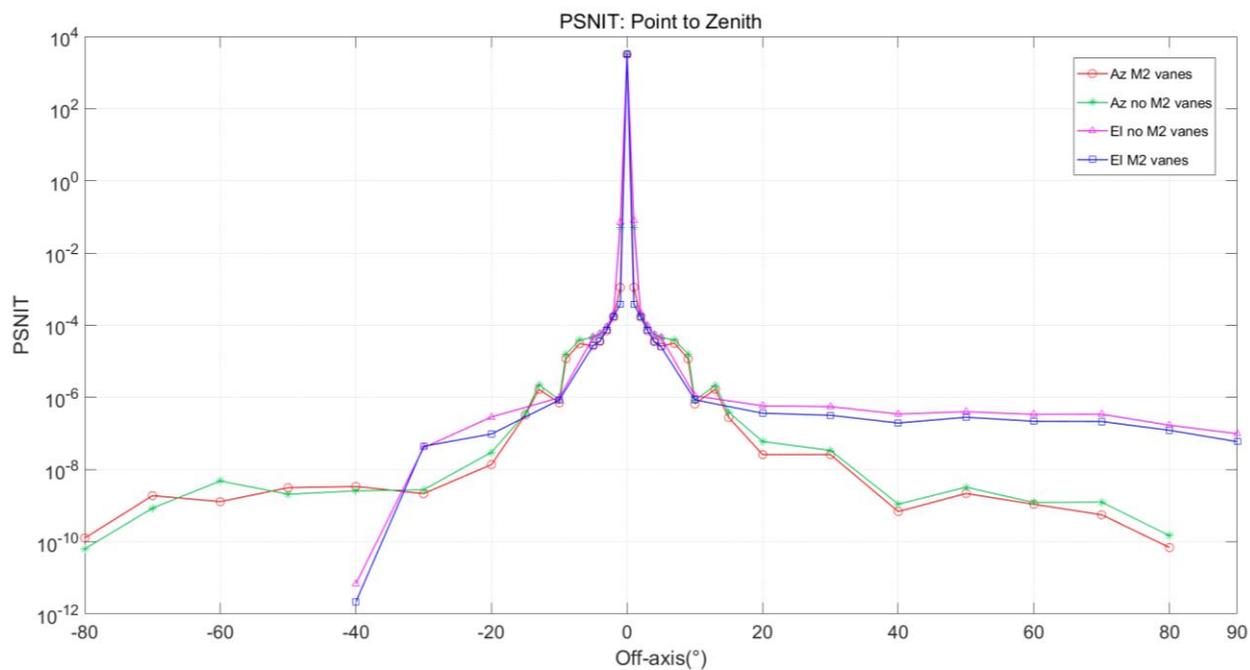

Figure 13 PSNIT with/ without vanes, telescope is pointed to zenith (Predicted reduction of intensity of scattered light).

Table 5 Rays number with, or without vanes

| Off-axis angle | no vane-AZ | vane-AZ | no vane-EL | vane-EL |
|---|---|---|---|---|
| 1 | 14684 | 13443 | 14604 | 11420 |
| 2 | 15631 | 13348 | 15650 | 11478 |
| 3 | 12499 | 10936 | 11098 | 9569 |
| 4 | 8766 | 7311 | 7430 | 6115 |
| 5 | 7593 | 6210 | 6438 | 5176 |
| 10 | 1974 | 1857 | 2244 | 2106 |
| 20 | 66 | 32 | 1300 | 1195 |
| 30 | 24 | 18 | 584 | 406 |
| 40 | 5 | 2 | 258 | 142 |
| 50 | 6 | 4 | 252 | 168 |
| 60 | 3 | 2 | 223 | 153 |
| 70 | 3 | 1 | 189 | 124 |
| 80 | 1 | 5 | 127 | 90 |
| 90 | - | - | 66 | 38 |
| Total | 61255 | 53169 | 60463 | 48180 |

## 6. CONCLUSIONS

As one of the campaigns of stray light control for Xinglong 2.16-m Telescope, a detail stray light analysis of the telescope has been completed, which is important to identify the stray light performance. The two complex stray light simulations, "point to 60°" and "point to zenith", containing simplified dome and telescope have been modeling. Identification of critical objects and illuminated objects, calculation of point source normalized irradiance transmittance curves and description of stray light contributors have been presented using TracePro at different off-axis angles. In the future, a number of new instruments including spectrograph, Tip-tilt and photopolarimeter will be available to the 2.16-m telescope, the stray light analysis of the 2.16-m telescope could provide an overview and significant specification for instrumentation performance improving.

## ACKNOWLEDGEMENTS

Work described in this paper is supported by National Natural Science Foundation of China, NSFC (Grant No. 11703043 and No. U1831209).

## REFERENCES

[1] Zhou Fan, et al. "The Xinglong 2.16-m Telescope: Current Instruments and Scientific Projects," PASP,128 (2016)
[2] Dingqiang Su, 2.16-m telescope work collections, *Science and technology of China press*, Beijing, (2001).
[3] M. Bass, Handbook of optics. *McGraw-Hill New York*, 2010.
[4] Eric.C. Fest, Engineers. Stray light analysis and control. *SPIE Press Bellingham*, 2013.
[5] F.O. Bartell, E. Dereniak, W. Wolfe. The theory and measurement of bidirectional reflectance distribution function (BRDF) and bidirectional transmittance distribution function (BTDF). *1980 Huntsville Technical Symposium*, 1981: 154-160.
[6] A.W. Greynolds. Stray light computations: Has nothing changed since the 1970s? *Optical Engineering+ Applications*, 2007: 66750B-66750B-66711.
[7] M. Schaub, J. Schwiegerling, E. Fest, et al. Molded Optics: Design and Manufacture. *CRC press*, 2011.
[8] P. Bely. The design and construction of large optical telescopes. *Springer Science & Business Media*, 2003.